\documentclass[12pt,a4paper]{article}
\usepackage{amssymb}
\usepackage{graphicx}
\newlength{\tabcolseporig}
\begin{document}
\null
\begin{flushright}
\begin{tabular}{r}
UWThPh-1997-14\\
DFTT 31/97\\
hep-ph/9705436\\
May 1997
\end{tabular}
\end{flushright}
\begin{center}
\Large \bfseries
NEUTRINO MASSES AND MIXING FROM NEUTRINO 
OSCILLATION EXPERIMENTS\footnote{
Talk presented by S.M. Bilenky at the
\emph{Fourth International Solar Neutrino Conference},
Heidelberg,
Germany,
April 8--11, 1997.}
\\[1cm]
\large \mdseries
S.M. Bilenky$^{\mathrm{(a)}}$,
C. Giunti$^{\mathrm{(b)}}$
and
W. Grimus$^{\mathrm{(c)}}$
\\[0.5cm]
\itshape
\normalsize
\setlength{\tabcolsep}{0pt}
\begin{tabular}{rl}
\large
$^{\mathrm{(a)}}$
&
\normalsize
Joint Institute for Nuclear Research, Dubna, Russia, and
\tabularnewline[0pt]
&
Technion, Physics Department, 32000 Haifa, Israel
\\[0.3cm]
\large
$^{\mathrm{(b)}}$
&
\normalsize
INFN, Sezione di Torino, 
and
Dipartimento di Fisica Teorica,
\tabularnewline[0pt]
&
Universit\`a di Torino,
Via P. Giuria 1, I--10125 Torino, Italy
\\[0.3cm]
\large
$^{\mathrm{(c)}}$
&
\normalsize
Institute for Theoretical Physics, University of Vienna,
\tabularnewline[0pt]
&
Boltzmanngasse 5, A--1090 Vienna, Austria
\end{tabular}
\setlength{\tabcolsep}{\tabcolseporig}
\\
\vspace*{1cm}
\upshape
\large
Abstract
\\[0.5cm]
\normalsize
\begin{minipage}[t]{0.9\textwidth}
What information about the neutrino 
mass spectrum and
mixing matrix
can be inferred from the existing 
neutrino oscillation data?
We discuss here the answer to this question
in the case of mixing of three and four neutrinos.
We present the
constraints on the effective
Majorana mass
$\langle{m}\rangle$
that can be obtained from
the results of reactor neutrino oscillation experiments
and
from atmospheric neutrino data.
We discuss
the bounds on
the oscillation probabilities in long-baseline
neutrino oscillation
experiments that follow from the results of short-baseline
experiments.
Some remarks on a model-independent approach
to the solar neutrino problem are also made.
\end{minipage}
\end{center}

\newpage
\renewcommand{\arraystretch}{1.5}

\section{Introduction}
\label{Introduction}

The problem of neutrino mass and mixing is the
most important problem of today's neutrino physics.
The hypothesis of neutrino mixing
was put forward by B. Pontecorvo in 1958
\cite{Pontecorvo}.
Many years later,
after the appearance of GUT models,
his idea became very popular.
Today the investigation of neutrino masses and mixing
is considered as one of the major ways of searching for new physics.

In accordance with the neutrino mixing hypothesis
(see, for example, Refs.\cite{BP78,BP87,CWKim}),
the fields
$\nu_{{\alpha}L}$
of flavour neutrinos
determined by 
the standard charged and neutral currents
\begin{equation}
j^{\mathrm{CC}}_{\rho}
=
2
\sum_{\alpha=e,\mu,\tau}
\bar\nu_{{\alpha}L}
\,
\gamma_{\rho}
\,
\ell_{{\alpha}L}
\;,
\qquad
j^{\mathrm{NC}}_{\rho}
=
\sum_{\alpha=e,\mu,\tau}
\bar\nu_{{\alpha}L}
\,
\gamma_{\rho}
\,
\nu_{{\alpha}L}
\;.
\label{ccnc}
\end{equation}
are  mixtures of the fields
of neutrinos with definite mass:
\begin{equation}
\nu_{{\alpha}L}
=
\sum_{k}
U_{{\alpha}k}
\,
\nu_{kL}
\;.
\label{mix1}
\end{equation}
Here $U$ is the unitary mixing matrix and
$\nu_{kL}$
is the field of the neutrino
with mass $m_k$.

Neutrino mixing can be quite different from the CKM
quark mixing.
Quarks are Dirac particles,
whereas for neutrinos with definite masses
there are two possibilities:
neutrinos can be Dirac
or truly neutral Majorana particles.
Dirac masses and mixing of neutrinos can be generated by the standard
Higgs mechanism.
Majorana masses and mixing require
a new mechanism of mass generation that does not conserve the total 
lepton charge.
Let us notice also that the number of massive neutrinos
in the general case of neutrino mixing can be more than 
the number of lepton flavours
which, according to LEP data,
is equal to three.
In this case,
in addition to Eq.(\ref{mix1}) we have
\begin{equation}
(\nu_{\bar{\alpha}R})^{c}
=
\sum_{k}
U_{\bar{\alpha}k}
\,
\nu_{kL}
\;,
\label{mix2}
\end{equation}
where
$\nu_{\bar{\alpha}R}$
is a right-handed sterile field
and
$
(\nu_{\bar{\alpha}R})^{c}
=
\mathcal{C} (\bar\nu_{\bar{\alpha}R})^{T}
$
($\mathcal{C}$ is the matrix of charge-conjugation).

From all existing data it follows that neutrino masses
(if any) are much smaller than the masses of all
the other fundamental fermions.
There is the very attractive see-saw mechanism \cite{see-saw}
of neutrino mass
generation that connects the smallness of neutrino masses
with the violation of lepton number
at a very large scale $M$ that characterizes the right-handed
Majorana mass term.
In this case,
for the neutrino masses
we have the relations
$
m_k
\sim
m_{\mathrm{F}k}^2 / M
$
($k=1,2,3$),
where
$ m_{\mathrm{F}k} $
is the mass of the up-quark or charged lepton
in the $k^{\mathrm{th}}$ generation
and
$ M \gg m_{\mathrm{F}k} $.
If the neutrino masses are generated with
the see-saw mechanism,
then
1)
the number of massive neutrinos is equal to three,
2)
massive neutrinos are Majorana particles,
3)
there is a hierarchy of neutrino masses:
\begin{equation}
m_1 \ll m_2 \ll m_3
\;.
\label{hierarchy}
\end{equation}

At present
there are three experimental indications in
favour of neutrino mixing.
The first
indication was obtained in solar neutrino
experiments: neutrino mixing
is the most natural explanation
of the deficit of solar $\nu_e$'s
observed in all solar neutrino experiments
(Homestake,
Kamiokande,
GALLEX
and SAGE
\cite{solar}).
The suppression of the solar $\nu_e$ flux
can be due to resonant MSW transitions with
a neutrino mass-squared difference
$ \sim 10^{-5} \, \mathrm{eV}^2 $.

The second indication in favour of neutrino oscillations
was obtained in the
Kamiokande, IMB and Soudan
atmospheric neutrino experiments
\cite{atmospheric}.
The observed deficit
of muon neutrinos can be explained by neutrino oscillations
with a mass-squared difference
$ \sim 10^{-2} \, \mathrm{eV}^2 $.

The third indication in favour of neutrino
mixing was obtained in the LSND experiment \cite{LSND}.
The observed number of $\bar\nu_e$ events can be explained by
$\bar\nu_\mu\to\bar\nu_e$ oscillations with
a mass-squared difference
$ \sim 1 \, \mathrm{eV}^2 $.

On the other hand,
no indication in favour of neutrino masses and mixing
was found in
numerous reactor and accelerator oscillation experiments 
(see the reviews in Ref.\cite{Boehm-Vannucci}),
in the experiments on the precise measurement of 
the high energy part of the beta-decay
spectrum of tritium
(see Ref.\cite{Breviews})
and
in the experiments
on the search for neutrinoless double-beta decay
(see Ref.\cite{BBreviews}).

We will address here the following question:
what information about neutrino mixing
and the neutrino mass spectrum can be obtained
from the existing data.
Some predictions
for the future experiments will also be discussed.

Let us start with
neutrino oscillations 
in short-baseline (SBL) experiments.
We will consider the general case of $n$ neutrinos 
with masses 
\begin{equation}
m_1 < m_2 < \ldots < m_{r-1}
\ll
m_r < \ldots < m_n
\label{masses}
\end{equation}
and we will assume
that only the largest mass square difference
$ \Delta{m}^2 \equiv m^2_n - m^2_1 $
is relevant for SBL oscillations
\cite{BBGK,BGKP,BGG96,BGG97}:
\begin{equation}
\frac{\Delta{m}^{2}L}{2p}
\gtrsim
1
\;,
\quad
\frac{\Delta{m}^{2}_{k1}L}{2p}
\ll
1
\; \mbox{for} \;
k < r
\quad \mbox{and} \quad
\frac{\Delta{m}^{2}_{nk} L}{2p}
\ll
1
\; \mbox{for} \;
k \geq r
\;,
\label{ass}
\end{equation}
where
$ \Delta{m}^{2}_{kj} \equiv m^2_k - m^2_j $,
$L$ is the
distance between
the neutrino source and detector
and $p$ is the neutrino momentum.

Using the unitarity of the mixing matrix,
for the amplitude of
$\nu_\alpha\to\nu_\beta$
transitions
we have
\begin{equation}
|\mathcal{A}_{\nu_\alpha\to\nu_\beta}|
=
\left|
\delta_{\alpha\beta}
+
\left(
\sum_{k=r}^{n}
U_{{\beta}k}
\,
U_{{\alpha}k}^{*}
\right)
\left[
\exp\!\left(
- i \frac{ \Delta{m}^{2} L }{ 2 p }
\right)
-
1
\right]
\right|
\;.
\label{amplitude}
\end{equation}
The probability of
$\nu_\alpha\to\nu_\beta$
transitions with $\alpha\neq\beta$
is given by
\begin{equation}
P_{\nu_{\alpha}\rightarrow\nu_{\beta}}
=
\frac{1}{2}
\,
A_{\alpha;\beta}
\left( 1 - \cos \frac{ \Delta{m}^{2} L }{ 2 p } \right)
\;,
\label{transition}
\end{equation}
with the oscillation amplitude
\begin{equation}
A_{\alpha;\beta}
=
4 \left| \sum_{k=r}^{n} U_{{\beta}k} \, U_{{\alpha}k}^{*} \right|^2
=
4 \left| \sum_{k=1}^{r-1} U_{{\beta}k} \, U_{{\alpha}k}^{*} \right|^2
\;.
\label{AA}
\end{equation}
For the survival probability of $\nu_{\alpha}$,
from Eqs.(\ref{transition}) and (\ref{AA})
we find
\begin{equation}
P_{\nu_{\alpha}\rightarrow\nu_{\alpha}}
=
1 - \sum_{\beta\neq\alpha}
P_{\nu_\alpha\to\nu_\beta}
=
1 - \frac{1}{2}
\,
B_{\alpha;\alpha}
\left(1 - \cos \frac{ \Delta{m}^{2} L }{ 2 p } \right)
\;,
\label{survival}
\end{equation}
where
\begin{equation}
\begin{array}{rcl} \displaystyle
B_{\alpha;\alpha}
=
\sum_{\beta\neq\alpha}
A_{\alpha;\beta}
\null & \null \displaystyle
=
\null & \null \displaystyle
4
\left( \sum_{k=r}^{n} |U_{{\alpha}k}|^2 \right)
\left( 1 - \sum_{k=r}^{n} |U_{{\alpha}k}|^2 \right)
\\ \displaystyle
\null & \null \displaystyle
=
\null & \null \displaystyle
4
\left( \sum_{k=1}^{r-1} |U_{{\alpha}k}|^2 \right)
\left( 1 - \sum_{k=1}^{r-1} |U_{{\alpha}k}|^2 \right)
\;.
\end{array}
\label{BB}
\end{equation}
It is obvious from Eqs.(\ref{transition})--(\ref{BB})
that
$ 0 \leq A_{\alpha;\beta} \leq 1 $
and
$ 0 \leq B_{\alpha;\alpha} \leq 1 $.

The formulas (\ref{transition}) and (\ref{survival})
have the form of the standard expressions
for the transition probabilities in the case of mixing of
two neutrinos
(see, for example, Refs.\cite{BP78,BP87,CWKim}).
Therefore,
if we identify
$A_{\alpha;\beta}$
or
$B_{\alpha;\alpha}$
with
$\sin^{2}2\theta$
($\theta$ is the mixing angle in the two-neutrino case),
we can use the results of the standard analyses of
the neutrino oscillation data.

\section{Three massive neutrinos}
\label{Three massive neutrinos}

Let us consider first the case of
three massive neutrinos
with the mass hierarchy (\ref{hierarchy}),
assuming that
$\Delta{m}^2_{21}$
is relevant for the suppression
of the flux of solar $\nu_e$'s
\cite{BBGK}.
In this case
$n=r=3$,
SBL oscillations
depend on
$\Delta{m}^{2}$,
$|U_{e3}|^2$,
$|U_{\mu3}|^2$
(the unitarity of $U$ implies that
$
|U_{\tau3}|^2
=
1
-
|U_{e3}|^2
-
|U_{\mu3}|^2
$),
and the oscillation amplitudes are given by
\begin{equation}
A_{\alpha;\beta}
=
4
\,
|U_{{\alpha}3}|^2
\,
|U_{{\beta}3}|^2
\;,
\qquad
B_{\alpha;\alpha}
=
4
\,
|U_{{\alpha}3}|^2
\left(
1
-
|U_{{\alpha}3}|^2
\right)
\;.
\label{AABB}
\end{equation}
In this section we do not consider
the atmospheric neutrino anomaly,
whose explanation,
together with the explanations of
the other indications in favour of neutrino oscillations,
requires at least four massive neutrinos
(see Section \ref{Four massive neutrinos}).

The reactor $\bar\nu_e$
and accelerator
$
\stackrel{\makebox[0pt][l]
{$\hskip-3pt\scriptscriptstyle(-)$}}{\nu_{\mu}}
$
disappearance experiments did not find
any positive indication in favour of neutrino
oscillations.
From the exclusion plots obtained from
the data of these experiments, 
at any fixed value of $\Delta{m}^2$
we have the following upper bound for
the oscillation amplitudes:
\begin{equation}
B_{\alpha;\alpha}
\leq
B_{\alpha;\alpha}^{0}
\qquad
(\alpha=e,\mu)
\;.
\label{BB0}
\end{equation}
The exclusion plots obtained by the
Bugey \cite{Bugey95}
$\bar\nu_e$
disappearance experiment
and by the
CDHS and CCFR \cite{CDHS84-CCFR84}
$
\stackrel{\makebox[0pt][l]
{$\hskip-3pt\scriptscriptstyle(-)$}}{\nu_{\mu}}
$
disappearance experiments
imply that
the amplitudes 
$B_{e;e}^{0}$
and
$B_{\mu;\mu}^{0}$
are small for any value of
$\Delta{m}^2$
in the wide interval
\begin{equation}
10^{-1}
\lesssim
\Delta{m}^2
\lesssim
10^{3} \, \mathrm{eV}^2
\;.
\label{interval}
\end{equation}
This means that the parameters
$|U_{e3}|^2$
and
$|U_{\mu3}|^2$
can be small or large (close to one, see Eq.(\ref{AABB})):
\begin{equation}
|U_{{\alpha}3}|^2
\leq
a^0_\alpha
\qquad \mbox{or} \qquad
|U_{{\alpha}3}|^2
\geq
1 - a^0_\alpha
\qquad
(\alpha=e,\mu)
\;,
\label{uu}
\end{equation}
where
\begin{equation}
a^0_\alpha
=
\frac{1}{2}
\left(
1
-
\sqrt{ 1 - B_{\alpha;\alpha}^{0} }
\right)
\;.
\label{a0}
\end{equation}
The quantity
$a^{0}_{e}$
is small
($ a^{0}_e \lesssim 4 \times 10^{-2} $)
for any value of
$\Delta{m}^{2}$
in the range (\ref{interval})
and
$a^{0}_{\mu}$
is small for
$
\Delta{m}^{2} \gtrsim 0.3 \, \mathrm{eV}^2
$
($ a^{0}_\mu \lesssim 10^{-1} $)
(see Ref.\cite{BBGK}).

From the results of
the solar neutrino experiments it follows
that only small values of
$|U_{e3}|^2$
are allowed.
In fact,
in the case of a neutrino mass hierarchy
that we are considering,
the probability of solar neutrinos to survive
is given by
(see Ref.\cite{SS92})
\begin{equation}
P_{\nu_e\to\nu_e}^{\mathrm{sun}}(E)
=
\left(
1
-
|U_{e3}|^2
\right)^2
P_{\nu_e\to\nu_e}^{(1,2)}(E)
+
|U_{e3}|^4
\;,
\label{solar}
\end{equation}
where
$ P_{\nu_e\to\nu_e}^{(1,2)}(E) $
is the $\nu_{e}$ survival probability 
due to the mixing between
the first and the second generations
and
$E$ is the neutrino energy.
Eq.(\ref{solar})
implies that
$
P_{\nu_e\to\nu_e}^{\mathrm{sun}}
\geq
|U_{e3}|^4
$.
If 
$ |U_{e3}|^2 \geq 1 - a^0_e $,
we have
$
P_{\nu_e\to\nu_e}^{\mathrm{sun}}
\geq
0.92
$
at all neutrino energies,
which is a bound that is not compatible
with the solar neutrino data.

Thus,
we come to the conclusion that  only two 
schemes are possible:
\begin{eqnarray}
(\mathrm{I})
\null & \null \qquad \null & \null
|U_{e3}|^2 \leq a^0_e
\;, \quad
|U_{\mu3}|^2 \leq a^0_\mu
\;,
\label{I}
\\
(\mathrm{II})
\null & \null \qquad \null & \null
|U_{e3}|^2 \leq a^0_e
\;, \quad
|U_{\mu3}|^2 \geq 1 - a^0_\mu
\;.
\label{II}
\end{eqnarray}

The amplitudes of
$
\stackrel{\makebox[0pt][l]
{$\hskip-3pt\scriptscriptstyle(-)$}}{\nu_{\mu}}
\to\stackrel{\makebox[0pt][l]
{$\hskip-3pt\scriptscriptstyle(-)$}}{\nu_{e}}
$
transitions
in the case of scheme I and 
$
\stackrel{\makebox[0pt][l]
{$\hskip-3pt\scriptscriptstyle(-)$}}{\nu_{e}}
\to\stackrel{\makebox[0pt][l]
{$\hskip-3pt\scriptscriptstyle(-)$}}{\nu_{\tau}}
$
transitions in case of scheme II
have upper bounds bilinear in
the small quantities
$a^0_e$,
$a^0_\mu$:
\begin{eqnarray}
(\mathrm{I})
\null & \null \qquad \null & \null
A_{\mu;e} \leq 4 \, a^0_e \, a^0_\mu
\;,
\label{amuel}
\\
(\mathrm{II})
\null & \null \qquad \null & \null
A_{e;\tau} \leq 4 \, a^0_e \, a^0_\mu
\;.
\label{aelta}
\end{eqnarray}
On the other hand,
the upper bound for the amplitude of 
$
\stackrel{\makebox[0pt][l]
{$\hskip-3pt\scriptscriptstyle(-)$}}{\nu_{\mu}}
\to\stackrel{\makebox[0pt][l]
{$\hskip-3pt\scriptscriptstyle(-)$}}{\nu_{\tau}}
$
transitions in both schemes is only linear in the small
quantity $a^0_\mu$:
$
A_{\mu;\tau} \leq 4 \, a^0_\mu
$.

\begin{figure}[p]
\begin{center}
\includegraphics[bb=30 55 550 790,width=0.9\textwidth]{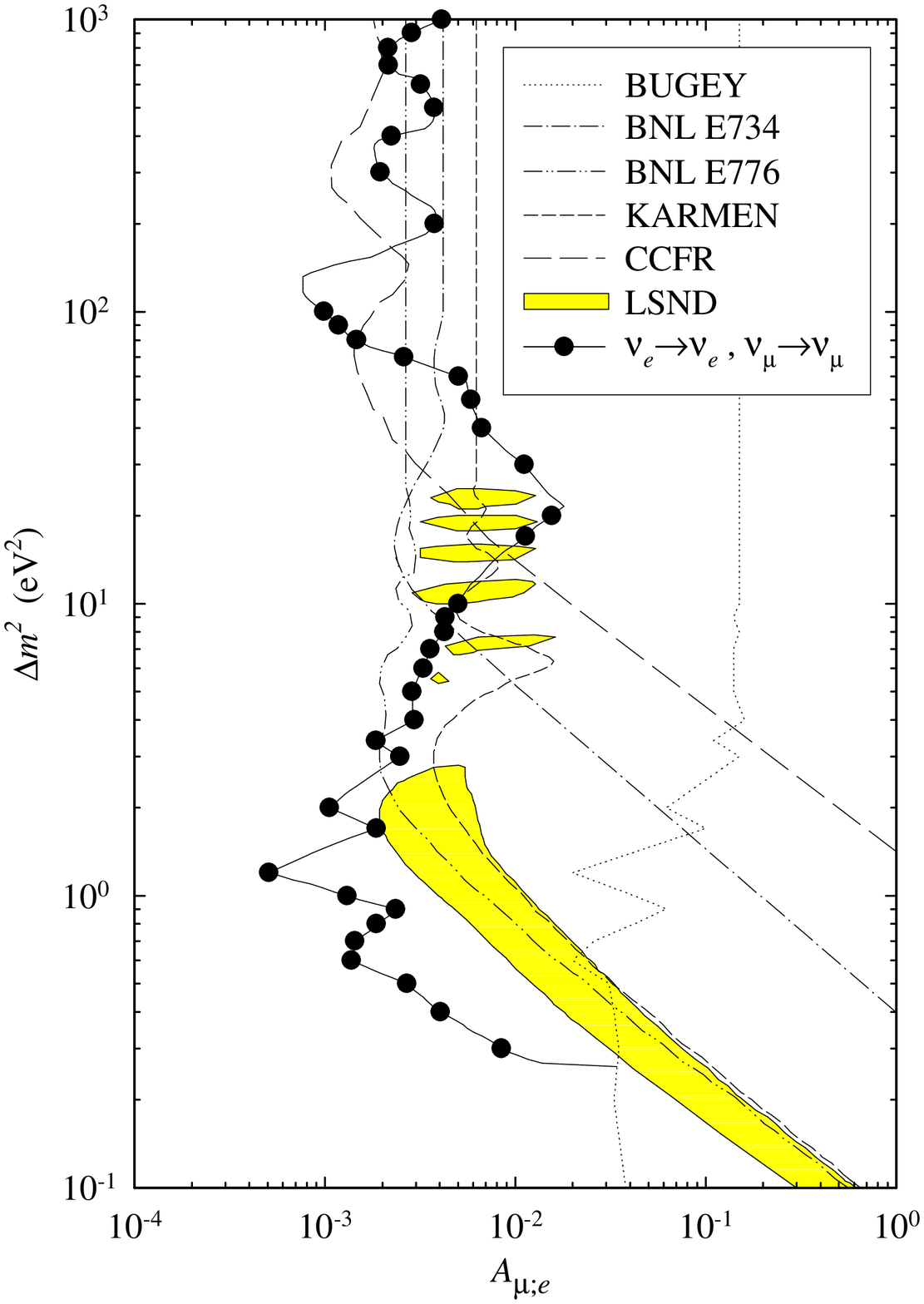}
\\[0.5cm]
Figure \ref{fig1}
\end{center}
\refstepcounter{figure}
\label{fig1}
\end{figure}

The inequality (\ref{amuel}) implies that
$
\stackrel{\makebox[0pt][l]
{$\hskip-3pt\scriptscriptstyle(-)$}}{\nu_{\mu}}
\to\stackrel{\makebox[0pt][l]
{$\hskip-3pt\scriptscriptstyle(-)$}}{\nu_{e}}
$
transition are strongly suppressed.
Is this 
inequality compatible with the results of the LSND experiment
in which indications in favour of $\bar\nu_\mu\to\bar\nu_e$
oscillations were found?
This question was considered in Ref.\cite{BBGK}.
The upper bound obtained
with the help of Eq.(\ref{amuel})
from the 90\% CL exclusion plots of the
Bugey \cite{Bugey95}
$\bar\nu_e$
disappearance experiment
and of the
CDHS and CCFR \cite{CDHS84-CCFR84}
$
\stackrel{\makebox[0pt][l]
{$\hskip-3pt\scriptscriptstyle(-)$}}{\nu_{\mu}}
$
disappearance experiments
is represented in Fig.\ref{fig1}
by the curve passing trough the circles.
The shadowed regions in Fig.\ref{fig1}
are allowed by LSND at 90\% CL.
Also shown are
are the 90\% CL exclusion curves found in the
BNL E734,
BNL E776,
KARMEN
and
CCFR
\cite{BNLE734-BNLE776-KARMEN-CCFR96}
$
\stackrel{\makebox[0pt][l]
{$\hskip-3pt\scriptscriptstyle(-)$}}{\nu_{\mu}}
\to\stackrel{\makebox[0pt][l]
{$\hskip-3pt\scriptscriptstyle(-)$}}{\nu_{e}}
$
appearance experiments
and in the Bugey experiment.
It is seen from Fig.\ref{fig1} that
the bounds that
were obtained from direct experiments on the search for
$
\stackrel{\makebox[0pt][l]
{$\hskip-3pt\scriptscriptstyle(-)$}}{\nu_{\mu}}
\to\stackrel{\makebox[0pt][l]
{$\hskip-3pt\scriptscriptstyle(-)$}}{\nu_{e}}
$
oscillations and the bound
(\ref{amuel})
obtained in the framework of scheme I
are not compatible
with the allowed regions of the LSND experiment
\cite{BBGK}.

Therefore,
we come to the conclusion that the scheme with
a hierarchy of neutrino masses and couplings
between generations (scheme I)
is not favoured by the existing experimental data.
A confirmation of the LSND
$\bar\nu_\mu\to\bar\nu_e$
signal would mean that
neutrino mixing
in the case of three massive neutrinos
is quite different from
quark mixing:
there is no hierarchy of couplings
and $\nu_\mu$
(not $\nu_\tau$)
is the ``heaviest'' neutrino.

\section{Neutrinoless double-beta decay}
\label{Neutrinoless double-beta decay}

We will discuss now the
limitations on the effective Majorana mass
\begin{equation}
\langle{m}\rangle
=
\sum_{k}
U_{ek}^2 \, m_k
\label{Majorana}
\end{equation}
that can be obtained from the neutrino oscillation data
in the framework of the 3-neutrino mass scheme
(\ref{hierarchy})
\cite{PS94,BBGK,BGM}.
As it is well known,
the effective mass $\langle{m}\rangle$
characterizes the contribution of Majorana neutrino 
masses and mixing to the matrix element
of neutrinoless double-beta decay
($(\beta\beta)_{0\nu}$)
in the case of a left-handed
interaction
(see, for example, Refs.\cite{BP87,CWKim}).

\begin{figure}[p]
\begin{center}
\includegraphics[bb=30 55 550 790,width=0.9\textwidth]{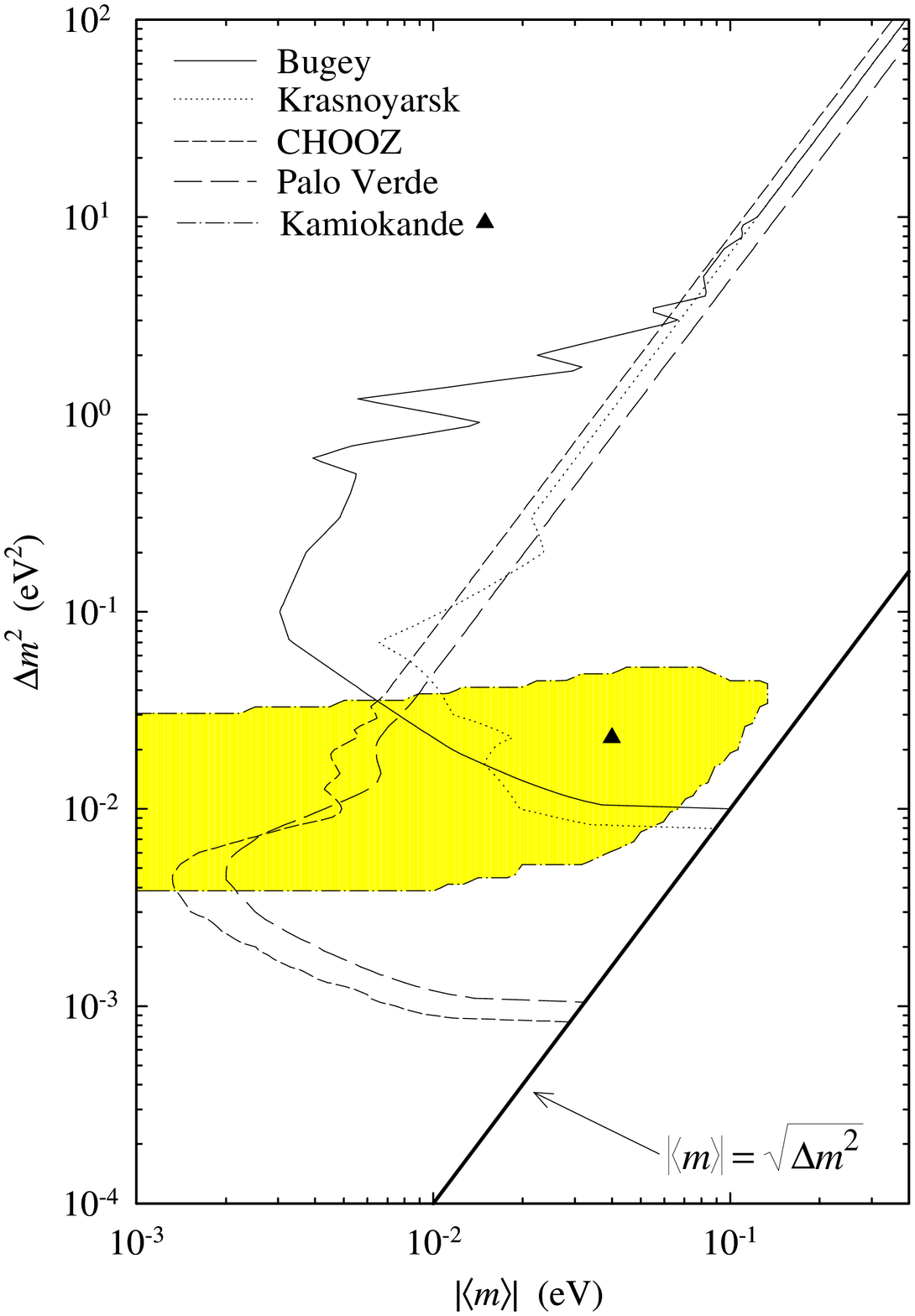}
\\[0.5cm]
Figure \ref{fig2}
\end{center}
\refstepcounter{figure}
\label{fig2}
\end{figure}

In the case of
three massive neutrinos
with the mass hierarchy (\ref{hierarchy})
and
$\Delta{m}^2_{21}$
relevant for the oscillations
of solar neutrinos,
from Eq.(\ref{Majorana})
and the constraint
$ |U_{e3}|^2 \leq a^0_e $
we have
\begin{equation}
|\langle{m}\rangle|
\simeq
|U_{e3}|^2 \, \sqrt{\Delta{m}^2}
\leq
a^0_e \, \sqrt{\Delta{m}^2}
\;.
\label{mmax}
\end{equation}
The solid line in Fig.\ref{fig2}
depicts this upper bound with $a^0_e$
obtained 
from the results of
the Bugey \cite{Bugey95}
and Krasnoyarsk \cite{Krasnoyarsk}
reactor experiments.
The straight line represents
the unitarity bound
$
|\langle{m}\rangle|
\leq
\sqrt{\Delta{m}^2}
$.
The dashed lines where obtained from
the sensitivity plots of the
CHOOZ and Palo Verde
\cite{CHOOZ-PaloVerde}
long-baseline (LBL) reactor 
experiments.
The shadowed region enclosed by the dash-dotted line
is allowed at 90\% CL
by the fit \cite{BGM}
of the Kamiokande
atmospheric neutrino data
(the black triangle corresponds to
the best value of the parameters).
In order to include in the figure this allowed region
and the sensitivity curves
of the CHOOZ and Palo Verde
experiments,
we considered $\Delta{m}^2$ in the wide range
$
10^{-4}
\leq
\Delta{m}^2
\leq
10^{2}
\, \mathrm{eV}^2
$.

As it is seen from Fig.\ref{fig2},
if
$ \Delta{m}^2 \lesssim 10^{-1} \, \mathrm{eV}^2 $
we have
$ |\langle{m}\rangle| \lesssim 10^{-1} \, \mathrm{eV}^2 $.
If
$ \Delta{m}^2 \lesssim 10^{2} \, \mathrm{eV}^2 $
we have
$ |\langle{m}\rangle| \lesssim 4 \times 10^{-1} \, \mathrm{eV}^2 $.
The results of the LSND experiment indicate that
$ 0.3 \lesssim \Delta{m}^2 \lesssim 3 \, \mathrm{eV}^2 $;
in this case
$ |\langle{m}\rangle| \lesssim 7 \times 10^{-2} \, \mathrm{eV}^2 $.
Finally,
the Kamiokande atmospheric neutrino data imply that
$ |\langle{m}\rangle| \lesssim 7 \times 10^{-2} \, \mathrm{eV}^2 $.

As it is well known,
the upper bounds on
$|\langle{m}\rangle|$
obtained from
experimental data of
$(\beta\beta)_{0\nu}$ decay
experiments
depend on the results of the calculation of nuclear
matrix elements.
The most stringent limit
\cite{Heidelberg-Moscow,Faessler96}
is given by the results of the
$^{76}$Ge experiments:
$
|\langle{m}\rangle|
<
( 0.6 - 1.6 ) \, \mathrm{eV}
$.
A big progress in searching for
$(\beta\beta)_{0\nu}$ decay
is expected in near future:
several collaborations plan to reach the sensitivity
$
|\langle{m}\rangle|
\simeq
( 0.1 - 0.3 ) \, \mathrm{eV}
$
\cite{Heidelberg-Moscow,BBfuture}.

Let us stress that the bound
(\ref{mmax})
is valid only in the case
of the neutrino mass hierarchy
(\ref{hierarchy}).
In the case of the inverted mass hierarchy
($n=3$ and $r=2$ in Eq.(\ref{masses}))
\cite{PS94,inverted,BGKP}
$
m_1 \ll m_2 \lesssim m_3
$,
with
$\Delta{m}^2_{32}$
relevant for solar neutrino oscillations,
$|\langle{m}\rangle|$
is limited only by the unitarity bound
$ |\langle{m}\rangle| \leq m_3 $.

Therefore,
we come to the conclusion that
the observation of neutrinoless
double-beta decay  
could allow to obtain
important information about
the spectrum of masses of Majorana neutrinos
\cite{PS94,BBGK}.

\section{Four massive neutrinos}
\label{Four massive neutrinos}

We will consider now 
the schemes with mixing of four massive neutrinos
which have
three different scales of
mass-squared differences
that correspond to
all existing indications in favour of neutrino
mixing
\cite{four}.
There are six possible schemes of
such type.
If the
results of solar, atmospheric, LSND
and all the other neutrino oscillation experiments
are taken into account,
only two schemes with the following mass spectra
are preferable:
\begin{equation}
(\mbox{A})
\qquad
\underbrace{
\overbrace{m_1 < m_2}^{\mathrm{atm}}
\ll
\overbrace{m_3 < m_4}^{\mathrm{solar}}
}_{\mathrm{LSND}}
\qquad \mbox{and} \qquad
(\mbox{B})
\qquad
\underbrace{
\overbrace{m_1 < m_2}^{\mathrm{solar}}
\ll
\overbrace{m_3 < m_4}^{\mathrm{atm}}
}_{\mathrm{LSND}}
\;.
\label{AB}
\end{equation}
In scheme A,
$\Delta{m}^{2}_{21}$
is relevant
for oscillations of atmospheric and LBL neutrinos
and
$\Delta{m}^{2}_{43}$
is relevant
for solar neutrino oscillations.
In scheme B,
the roles of
$\Delta{m}^{2}_{21}$
and
$\Delta{m}^{2}_{43}$
are reversed.

In order to see that
the results of neutrino oscillation experiments
indicate the neutrino spectra (\ref{AB}),
let us consider,
for example,
the neutrino spectra with one mass,
$m_4$,
separated from other three masses by a
gap of about 1 eV
($n=r=4$ in Eq.(\ref{masses})):
\begin{equation}
m_1 < m_2 < m_3 \ll m_4
\;.
\label{4hierarchy}
\end{equation}
In this case
SBL oscillations
depend on four parameters,
$\Delta{m}^{2}$,
$|U_{e4}|^2$,
$|U_{\mu4}|^2$,
$|U_{\tau4}|^2$,
and the oscillation probabilities are given
by Eqs.(\ref{transition}) and (\ref{survival})
with the oscillation amplitudes
\begin{equation}
A_{\alpha;\beta}
=
4
\,
|U_{{\alpha}4}|^2
\,
|U_{{\beta}4}|^2
\;,
\qquad
B_{\alpha;\alpha}
=
4
\,
|U_{{\alpha}4}|^2
\left(
1
-
|U_{{\alpha}4}|^2
\right)
\;.
\label{AABB4hierarchy}
\end{equation}
Using the same reasoning 
as in Section \ref{Three massive neutrinos}
and taking into account
the results of solar neutrino experiments,
we come to the conclusion that
$ |U_{e4}|^2 \leq a^0_e $.
Now we must also take into account
the atmospheric neutrino anomaly.
For the probability of atmospheric $\nu_\mu$'s to survive
we have the lower bound
$
P^{\mathrm{atm}}_{\nu_\mu\to\nu_\mu}
\geq
|U_{\mu4}|^4
$
\cite{BGG96}.
From this inequality we conclude that,
in order to explain the atmospheric neutrino anomaly
from
the two possibilities
for $|U_{\mu4}|^2$
given by the results
of SBL
$
\stackrel{\makebox[0pt][l]
{$\hskip-3pt\scriptscriptstyle(-)$}}{\nu}_{\mu}
$
disappearance experiments,
$ |U_{\mu4}|^2 \leq a^0_\mu $
and
$ |U_{\mu4}|^2 \geq 1 - a^0_\mu $,
we must choose the first one.
Therefore,
the solution of the solar and atmospheric neutrino problems
and the results of reactor and accelerator
disappearance experiments
lead to the constraints
$ |U_{e4}|^2 \leq a^0_e $,
$ |U_{\mu4}|^2 \leq a^0_\mu $
and
for the amplitude of
$
\stackrel{\makebox[0pt][l]
{$\hskip-3pt\scriptscriptstyle(-)$}}{\nu_{\mu}}
\to\stackrel{\makebox[0pt][l]
{$\hskip-3pt\scriptscriptstyle(-)$}}{\nu_{e}}
$
transitions
we have the same bound (\ref{amuel})
as in the 3-neutrino scheme I.
The experimental situation on
$
\stackrel{\makebox[0pt][l]
{$\hskip-3pt\scriptscriptstyle(-)$}}{\nu_{\mu}}
\to\stackrel{\makebox[0pt][l]
{$\hskip-3pt\scriptscriptstyle(-)$}}{\nu_{e}}
$
oscillations is
presented in Fig.\ref{fig1} and 
we conclude that the neutrino mass spectra
under consideration 
are not favoured by the experimental data.
By the same reasons,
the mass spectra with the lightest mass
$m_1$
separated from the other three masses by a
gap of about 1 eV
are also not favoured by the data.

We will consider now
the two schemes (\ref{AB}) with
the mass spectra A and B. 
For the transition amplitudes 
in SBL experiments,
from the general expressions
(\ref{AA}) and (\ref{BB})
in both schemes we have
\begin{eqnarray}
&&
A_{\alpha;\beta}
=
4 \left| \sum_{k=1,2} U_{{\beta}k} U_{{\alpha}k}^{*} \right|^2
=
4 \left| \sum_{k=3,4} U_{{\beta}k} U_{{\alpha}k}^{*} \right|^2
\;,
\label{AA4}
\\
&&
\begin{array}{rcl} \displaystyle
B_{\alpha;\alpha}
\null & \null \displaystyle
=
\null & \null \displaystyle
4
\left( \sum_{k=1,2} |U_{{\alpha}k}|^2 \right)
\left( 1 - \sum_{k=1,2} |U_{{\alpha}k}|^2 \right)
\\ \displaystyle
\null & \null \displaystyle
=
\null & \null \displaystyle
4
\left( \sum_{k=3,4} |U_{{\alpha}k}|^2 \right)
\left( 1 - \sum_{k=3,4} |U_{{\alpha}k}|^2 \right)
\;.
\end{array}
\label{BB4}
\end{eqnarray}
Let us define the parameters
\begin{equation}
c_{\alpha}
\equiv
\sum_{k=1,2} |U_{{\alpha}k}|^2
\qquad \mbox{and} \qquad
d_{\alpha}
\equiv
\sum_{k=3,4} |U_{{\alpha}k}|^2
\;.
\label{dcc}
\end{equation}
It is obvious that the unitarity of the mixing matrix
requires that
\begin{equation}
c_{\alpha} + d_{\alpha} = 1
\;.
\label{c+d}
\end{equation}
Taking into account the
results of the solar and atmospheric neutrino experiments
and the constraints from the reactor
and accelerator disappearance experiments,
in the schemes A and B we have
\cite{BGG96}
\begin{eqnarray}
(\mbox{A})
\null & \null \qquad \null & \null
c_{e} \leq a^{0}_{e}
\;,
\qquad
d_{\mu} \leq a^{0}_{\mu}
\;,
\label{cda}
\\
(\mbox{B})
\null & \null \qquad \null & \null
d_{e} \leq a^{0}_{e}
\;,
\qquad
c_{\mu} \leq a^{0}_{\mu}
\;.
\label{cdb}
\end{eqnarray}

The schemes A and B give different predictions
for the effective neutrino mass
$m_{\nu}(^3\mathrm{H})$
measured in tritium experiments and
for the effective Majorana mass
$\langle{m}\rangle$
that determines the matrix element
of neutrinoless double-beta decays.
In fact, we have
\begin{equation}
\begin{array}{lllll} \displaystyle
(\mbox{A})
\null & \null \displaystyle
\quad
\null & \null \displaystyle
m_{\nu}(^3\mathrm{H})
\simeq
d_e m_4
\simeq
m_4
\;,
\null & \null \displaystyle
\quad
\null & \null \displaystyle
\langle{m}\rangle
\simeq
\left|
\sum_{k=3,4}
U_{ek}^2
\right|
m_4
\leq
d_e m_4
\simeq
m_4
\;,
\\ \displaystyle
(\mbox{B})
\null & \null \displaystyle
\quad
\null & \null \displaystyle
m_{\nu}(^3\mathrm{H})
\simeq
d_e m_4
\leq
a^{0}_{e} m_4
\ll m_4
\;,
\null & \null \displaystyle
\quad
\null & \null \displaystyle
\langle{m}\rangle
\leq
d_e m_4
\leq
a^{0}_{e} m_4
\ll m_4
\;.
\end{array}
\label{ABmm}
\end{equation}
Thus, if scheme A is realized in nature
the tritium experiments
and the experiments on the search for neutrinoless double beta
decay have a good chance to reveal
the effects of the heaviest
neutrino mass $m_4$.

Let us consider now neutrino oscillations in
long-baseline experiments.
The probabilities of
$\nu_\alpha\to\nu_\beta$
transitions
in LBL experiments in the schemes A and B are given by
\begin{eqnarray}
&&
P^{(\mathrm{LBL,A})}_{\nu_\alpha\to\nu_\beta}
=
\left|
\sum_{k=1,2}
U_{{\beta}k}
\,
U_{{\alpha}k}^{*}
\,
\exp\!\left(
- i
\frac{ \Delta{m}^{2}_{k1} \, L }{ 2 \, p }
\right)
\right|^2
+
\left|
\sum_{k=3,4}
U_{{\beta}k}
\,
U_{{\alpha}k}^{*}
\right|^2
\;,
\label{plbla}
\\
&&
P^{(\mathrm{LBL,B})}_{\nu_\alpha\to\nu_\beta}
=
\left|
\sum_{k=1,2}
U_{{\beta}k}
\,
U_{{\alpha}k}^{*}
\right|^2
+
\left|
\sum_{k=3,4}
U_{{\beta}k}
\,
U_{{\alpha}k}^{*}
\,
\exp\!\left(
i
\frac{ \Delta{m}^{2}_{4k} \, L }{ 2 \, p }
\right)
\right|^2
\;.
\label{plblb}
\end{eqnarray}
These formulas have been obtained
taking into account the fact that in LBL experiments
$ \Delta{m}^{2}_{43} L / 2 p \ll 1 $
in scheme A
and
$ \Delta{m}^{2}_{21} L / 2 p \ll 1 $
in scheme B
and dropping the terms proportional to 
the cosines of phases much larger
than $2\pi$
($ \Delta{m}^{2}_{kj} L / 2 p \gg 2\pi $
for $k=3,4$ and $j=1,2$),
which do not contribute to the oscillation
probabilities averaged over the
neutrino energy spectrum.
The oscillation probabilities
for antineutrinos
are given by the same expressions
with the changes
$ U_{{\beta}k} \to U_{{\beta}k}^{*} $
and
$ U_{{\alpha}k}^{*} \to U_{{\alpha}k} $.
With the help of the Cauchy--Schwarz inequality,
from
Eqs.(\ref{plbla})
for the survival probability of
$
\stackrel{\makebox[0pt][l]
{$\hskip-3pt\scriptscriptstyle(-)$}}{\nu_{\alpha}}
$
and
the probability of
$
\stackrel{\makebox[0pt][l]
{$\hskip-3pt\scriptscriptstyle(-)$}}{\nu_{\alpha}}
\to\stackrel{\makebox[0pt][l]
{$\hskip-3pt\scriptscriptstyle(-)$}}{\nu_{\beta}}
$
transitions
in LBL experiments
in the scheme A
we have the following bounds:
\begin{eqnarray}
&&
d_{\alpha}^2
\leq
P^{(\mathrm{LBL})}_{\stackrel{\makebox[0pt][l]
{$\hskip-3pt\scriptscriptstyle(-)$}}{\nu_{\alpha}}
\to\stackrel{\makebox[0pt][l]
{$\hskip-3pt\scriptscriptstyle(-)$}}{\nu_{\alpha}}}
\leq
c_{\alpha}^2
+
d_{\alpha}^2
\;,
\label{paa}
\\
&&
\frac{1}{4}
\,
A_{\alpha;\beta}
\leq
P^{(\mathrm{LBL})}_{\stackrel{\makebox[0pt][l]
{$\hskip-3pt\scriptscriptstyle(-)$}}{\nu_{\alpha}}
\to\stackrel{\makebox[0pt][l]
{$\hskip-3pt\scriptscriptstyle(-)$}}{\nu_{\beta}}}
\leq
c_{\alpha}
\,
c_{\beta}
+
\frac{1}{4}
\,
A_{\alpha;\beta}
\;,
\label{pab1}
\end{eqnarray}
where
$A_{\alpha;\beta}$
is the amplitude of
$
\stackrel{\makebox[0pt][l]
{$\hskip-3pt\scriptscriptstyle(-)$}}{\nu_{\alpha}}
\to\stackrel{\makebox[0pt][l]
{$\hskip-3pt\scriptscriptstyle(-)$}}{\nu_{\beta}}
$
oscillations in SBL experiments
(see Eq.(\ref{AA})).
The corresponding bounds in the scheme B
can be obtained with the change
$ c_\alpha \leftrightarrows d_\alpha $.
It is clear from Eqs.(\ref{cda}) and (\ref{cdb})
that
the bounds for the oscillation probabilities
in LBL experiments are equal
in the schemes A and B.

From Eqs.(\ref{paa}) and (\ref{pab1}),
using the limits on the parameters
$c_e$
and
$A_{\mu;e}$
obtained from the results of
reactor and accelerator SBL oscillation
experiments,
it is possible to obtain rather strong constraints
on the probabilities of
$
\stackrel{\makebox[0pt][l]
{$\hskip-3pt\scriptscriptstyle(-)$}}{\nu_{e}}
\to\stackrel{\makebox[0pt][l]
{$\hskip-3pt\scriptscriptstyle(-)$}}{\nu_{e}}
$
and
$
\stackrel{\makebox[0pt][l]
{$\hskip-3pt\scriptscriptstyle(-)$}}{\nu_{\mu}}
\to\stackrel{\makebox[0pt][l]
{$\hskip-3pt\scriptscriptstyle(-)$}}{\nu_{e}}
$
transitions
in LBL experiments
\cite{BGG97}.

\begin{figure}[p]
\begin{center}
\includegraphics[bb=30 55 550 790,width=0.9\textwidth]{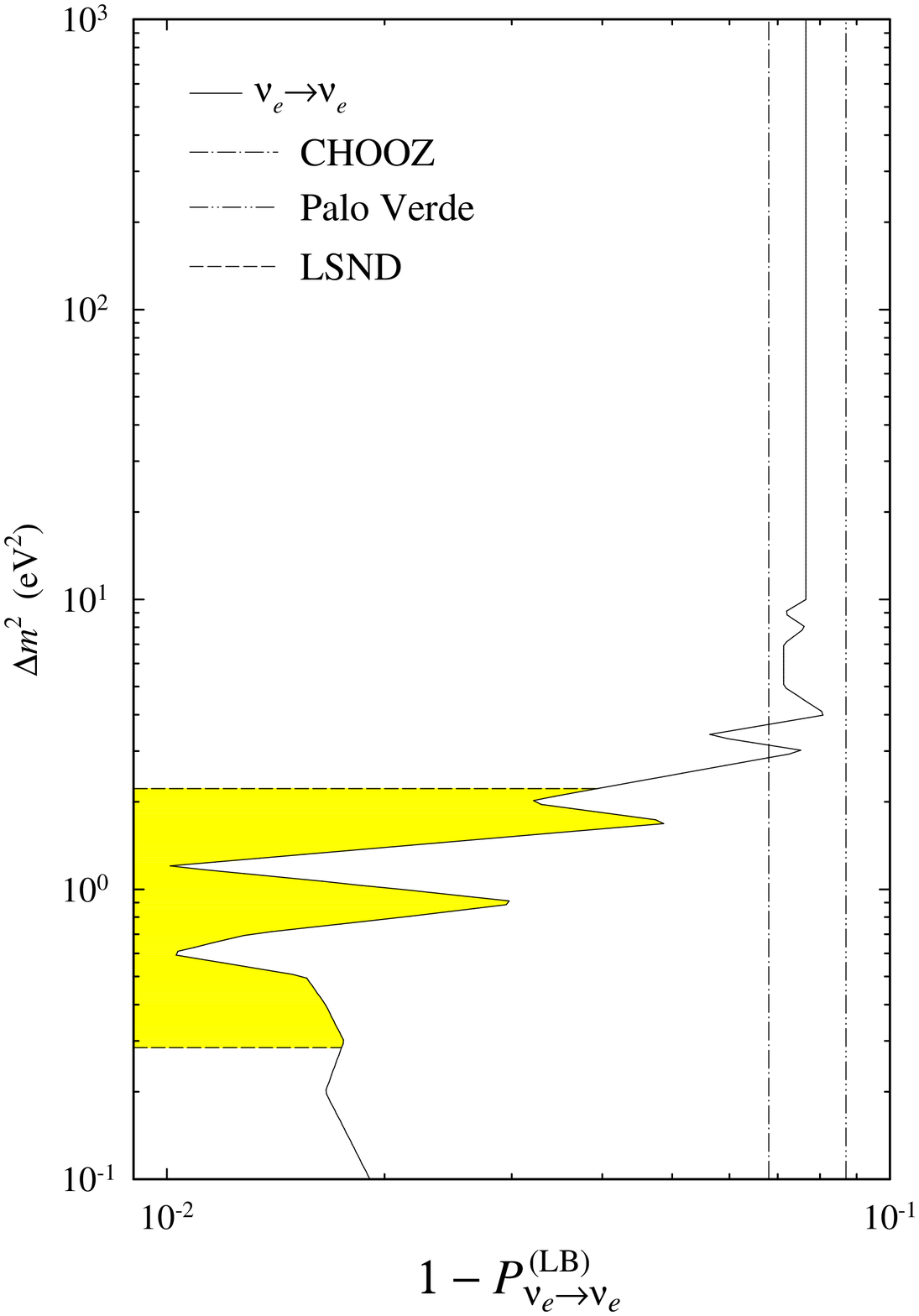}
\\[0.5cm]
Figure \ref{fig3}
\end{center}
\refstepcounter{figure}
\label{fig3}
\end{figure}

For the transition probability of
$
\stackrel{\makebox[0pt][l]
{$\hskip-3pt\scriptscriptstyle(-)$}}{\nu_{e}}
$
into all possible states, 
from Eqs.(\ref{cda}), (\ref{cdb}) and (\ref{paa})
we have the following bound
\begin{equation}
1
-
P^{(\mathrm{LBL})}_{\stackrel{\makebox[0pt][l]
{$\hskip-3pt\scriptscriptstyle(-)$}}{\nu_{e}}
\to\stackrel{\makebox[0pt][l]
{$\hskip-3pt\scriptscriptstyle(-)$}}{\nu_{e}}}
\leq
a^{0}_{e}
\left( 2 - a^{0}_{e} \right)
\;.
\label{1-pee}
\end{equation}
The curve corresponding
to this limit
obtained from the 90\% CL exclusion plot of the Bugey
\cite{Bugey95}
experiment is shown
in Fig.\ref{fig3}
(solid line).
The range of the SBL parameter
$\Delta{m}^2$
considered is
$
10^{-1}
\leq
\Delta{m}^2
\leq
10^{3}
\, \mathrm{eV}^2
$.
The dash-dotted and dash-dot-dotted vertical lines
depict
the minimal probability (sensitivities)
that will be reached by
CHOOZ and Palo Verde
long-baseline reactor neutrino experiments.
The shadowed region in Fig.\ref{fig3}
is allowed by the results of
LSND experiment.
Thus,
in the framework of the schemes A and B,
the CHOOZ experiment could
reveal LBL neutrino oscillations if
$ \Delta{m}^2 \gtrsim 4 \mathrm{eV}^2 $.

\begin{figure}[p]
\begin{center}
\includegraphics[bb=30 55 550 790,width=0.9\textwidth]{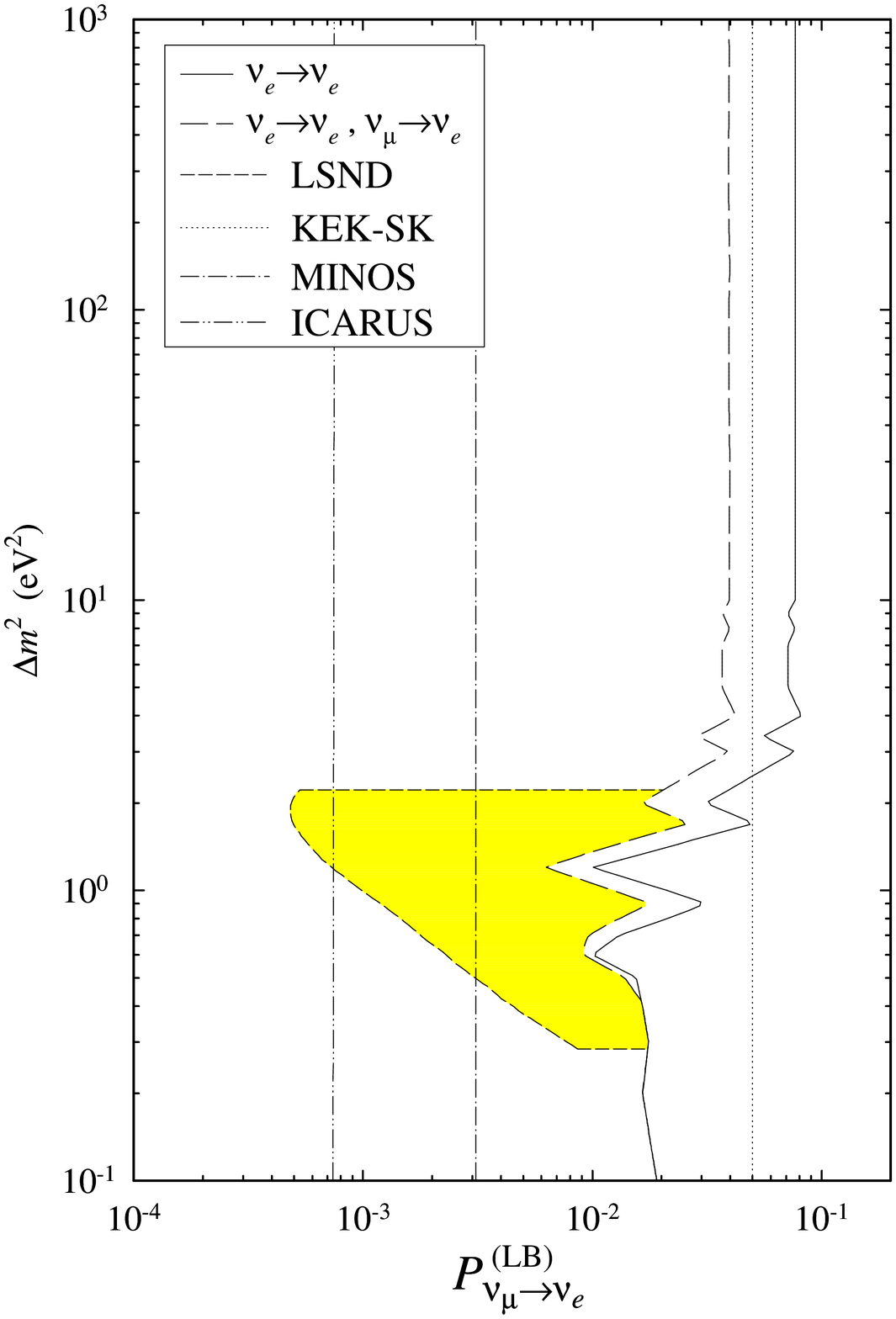}
\\[0.5cm]
Figure \ref{fig4}
\end{center}
\refstepcounter{figure}
\label{fig4}
\end{figure}

Let us consider now
$
\stackrel{\makebox[0pt][l]
{$\hskip-3pt\scriptscriptstyle(-)$}}{\nu_{\mu}}
\to\stackrel{\makebox[0pt][l]
{$\hskip-3pt\scriptscriptstyle(-)$}}{\nu_{e}}
$
transitions
in LBL experiments.
From Eqs.(\ref{cda}), (\ref{cdb}) and (\ref{pab1}),
we have the upper bound
\begin{equation}
P^{(\mathrm{LBL})}_{\stackrel{\makebox[0pt][l]
{$\hskip-3pt\scriptscriptstyle(-)$}}{\nu_{\mu}}
\to\stackrel{\makebox[0pt][l]
{$\hskip-3pt\scriptscriptstyle(-)$}}{\nu_{e}}}
\leq
a^{0}_{e}
+
\frac{1}{4}
\,
A_{\mu;e}^{0}
\;.
\label{pme1}
\end{equation}
where
$A_{\mu;e}^{0}$
is the upper bound for the amplitude of
$
\stackrel{\makebox[0pt][l]
{$\hskip-3pt\scriptscriptstyle(-)$}}{\nu_{\mu}}
\to\stackrel{\makebox[0pt][l]
{$\hskip-3pt\scriptscriptstyle(-)$}}{\nu_{e}}
$
transitions found in SBL experiments.
Another inequality can be obtained 
from Eq.(\ref{1-pee})
using the unitarity of the mixing matrix:
\begin{equation}
P^{(\mathrm{LBL})}_{\stackrel{\makebox[0pt][l]
{$\hskip-3pt\scriptscriptstyle(-)$}}{\nu_{\mu}}
\to\stackrel{\makebox[0pt][l]
{$\hskip-3pt\scriptscriptstyle(-)$}}{\nu_{e}}}
\leq
a^{0}_{e}
\left( 2 - a^{0}_{e} \right)
\;.
\label{pme2}
\end{equation}
The curves corresponding
to the limits (\ref{pme2})
(solid lines)
and (\ref{pme1})
(long-dashed line)
obtained from the 90\% CL exclusion plots of the Bugey
\cite{Bugey95}
experiment for
$a^{0}_{e}$
and
of the
BNL E734,
BNL E776
and
CCFR
\cite{BNLE734-BNLE776-KARMEN-CCFR96}
experiments
for
$A_{\mu;e}^{0}$
are shown
in Fig.\ref{fig4}.
Sensitivities
of the
KEK--SK, MINOS and ICARUS
\cite{KEKSK-MINOS-ICARUS}
LBL accelerator experiments
are represented in Fig.\ref{fig4} by
the dotted, dash-dotted and dash-dot-dotted vertical lines.
The
shadowed region is the region allowed
by the results of the LSND experiment.
As it is seen from Fig.\ref{fig4},
the sensitivities of the MINOS and ICARUS experiments
are much higher than the bounds that we have obtained.
The solid line in Fig.\ref{fig4} represents also
an upper bound for the
probablitiy of
$
\stackrel{\makebox[0pt][l]
{$\hskip-3pt\scriptscriptstyle(-)$}}{\nu_{e}}
\to\stackrel{\makebox[0pt][l]
{$\hskip-3pt\scriptscriptstyle(-)$}}{\nu_{\tau}}
$
transitions.

In the framework of schemes A and B
the results of SBL neutrino oscillation experiments do not put any
constraints
on the probabilities of
$
\stackrel{\makebox[0pt][l]
{$\hskip-3pt\scriptscriptstyle(-)$}}{\nu_{\mu}}
\to\stackrel{\makebox[0pt][l]
{$\hskip-3pt\scriptscriptstyle(-)$}}{\nu_{\mu}}
$
and
$
\stackrel{\makebox[0pt][l]
{$\hskip-3pt\scriptscriptstyle(-)$}}{\nu_{\mu}}
\to\stackrel{\makebox[0pt][l]
{$\hskip-3pt\scriptscriptstyle(-)$}}{\nu_{\tau}}
$
transitions in LBL experiments.
From our analysis it follows that 
$
\stackrel{\makebox[0pt][l]
{$\hskip-3pt\scriptscriptstyle(-)$}}{\nu_{\mu}}
\to\stackrel{\makebox[0pt][l]
{$\hskip-3pt\scriptscriptstyle(-)$}}{\nu_{\mu}}
$
and
$
\stackrel{\makebox[0pt][l]
{$\hskip-3pt\scriptscriptstyle(-)$}}{\nu_{\mu}}
\to\stackrel{\makebox[0pt][l]
{$\hskip-3pt\scriptscriptstyle(-)$}}{\nu_{\tau}}
$
are the preferable channels for
future LBL accelerator neutrino experiments.

\section{Solar neutrinos}
\label{Solar neutrinos}

In this Section
we will present a few remarks about solar neutrinos.
Most analyses of the data of solar neutrino experiments
are based on the Standard Solar Model
\cite{SSM}.
In spite of the great success of
the model,
it is very important to check
its predictions and to obtain
model-independent information about neutrino 
mixing
from solar neutrino experiments.

It was shown in Ref.\cite{BG} that
when the data of
the Super-Kamiokande (S-K)
\cite{SK}
and SNO
\cite{SNO}
experiments
will be available it will become possible:
1) to check whether there are transitions of solar $\nu_e$'s
into other states;
2) to measure the initial flux
of solar $^8$B $\nu_e$'s; 
3) to determine the probability of solar $\nu_e$'s to survive
directly from experimental data.

Important features of future solar neutrino experiments
will be the detection of solar neutrinos via 
CC and NC reactions
and the relatively large statistics of events.
In the S-K experiment solar neutrinos
are detected by the observation of the
elastic scattering (ES) process
$ \nu e^{-} \to \nu e^{-} $.
The spectrum of the recoil electrons
in this process can be 
written in the form
\begin{eqnarray}
n^{\mathrm{ES}}(T)
\null & \null = \null & \null
\int_{E_{\mathrm{m}}(T)}
\mathrm{d}E
\,
\left[
\frac
{ \mathrm{d} \sigma_{\nu_{e}e} }
{ \mathrm{d} T }
(E,T)
-
\frac
{ \mathrm{d} \sigma_{\nu_{\mu}e} }
{ \mathrm{d} T }
(E,T)
\right]
\phi_{\nu_{e}}(E)
\\
\null & \null & \null
+
\int_{E_{\mathrm{m}}(T)}
\mathrm{d}E
\,
\frac
{ \mathrm{d} \sigma_{\nu_{\mu}e} }
{ \mathrm{d} T }
(E,T)
\,
\phi_{\nu_{e}}^{0}(E)
\sum_{\beta=e,\mu,\tau}
P^{\mathrm{sun}}_{\nu_e\to\nu_\beta}(E)
\;.
\label{nes}
\end{eqnarray}
Here T is the kinetic energy of the recoil electron,
$
E_{\mathrm{m}}(T)
=
\frac{ T }{ 2 }
\left(
1
+
\sqrt{ 1 + \frac{ 2 m_{e} }{ T } }
\right)
$,
$
\frac
{ \mathrm{d} \sigma_{\nu_{\alpha}e} }
{ \mathrm{d} T }
(E,T)
$
is the differential cross section
of the process
$ \nu_{\alpha} e \to \nu_{\alpha} e $
($\alpha=e,\mu$),
$\phi_{\nu_{e}}(E)$
is the spectrum of solar $\nu_e$'s on the Earth
and
$\phi_{\nu_{e}}^{0}(E)$
is the spectrum of initial $^8$B neutrinos,
which can be written as
\begin{equation}
\phi_{\nu_{e}}^{0}(E)
=
\Phi_{\mathrm{B}}
\,
X(E)
\label{flux}
\end{equation}
were $X(E)$ is a known normalized function and
$\Phi_{\mathrm{B}}$
is the total flux.
We can rewrite the relation (\ref{nes})
in the form
\begin{equation}
\frac
{ \Sigma^{\mathrm{ES}}(T) }
{ X_{\nu_{\mu}e}(T) }
=
\left\langle
\sum_{\beta=e,\mu,\tau}
P^{\mathrm{sun}}_{\nu_{e}\to\nu_{\beta}}
\right\rangle_{{\hskip-2pt}T}
\Phi_{\mathrm{B}}
\;,
\label{pave}
\end{equation}
where
\begin{equation}
\Sigma^{\mathrm{ES}}(T)
\equiv
n^{\mathrm{ES}}(T)
-
\int_{E_{\mathrm{m}}(T)}
\mathrm{d} E
\left[
\frac
{ \mathrm{d} \sigma_{\nu_{e}e} }
{ \mathrm{d} T }
(E,T)
-
\frac
{ \mathrm{d} \sigma_{\nu_{\mu}e} }
{ \mathrm{d} T }
(E,T)
\right]
\phi_{\nu_{e}}(E)
\label{ses}
\end{equation}
and
\begin{equation}
X_{\nu_{\mu}e}(T)
\equiv
\int_{E_{\mathrm{m}}(T)}
\mathrm{d} E
\,
\frac{ \mathrm{d} \sigma_{\nu_{\mu}e} }
{ \mathrm{d} T }
(E,T)
\,
X(E)
\label{E533}
\end{equation}
is a known function.
The quantity
$
\left\langle
\sum_{\beta=e,\mu,\tau}
P^{\mathrm{sun}}_{\nu_{e}\to\nu_{\beta}}
\right\rangle_{{\hskip-2pt}T}
$
is the average over
$
\frac{ \mathrm{d} \sigma_{\nu_{\mu}e} }
{ \mathrm{d} T }
(E,T)
\,
X(E)
$
of
the total transition probability of
solar $\nu_e$'s into all possible active states.
If there are no transitions of solar neutrinos
into sterile states,
we have
$
\left\langle
\sum_{\beta=e,\mu,\tau}
P^{\mathrm{sun}}_{\nu_{e}\to\nu_{\beta}}
\right\rangle_{{\hskip-2pt}T}
=
1
$.

The function
$\Sigma^{\mathrm{ES}}(T)$
can be determined by combining the S-K
measurement of the spectrum 
of recoil electrons
with the measurement of the spectrum
$\phi_{\nu_{e}}(E)$
of $\nu_e$'s on the Earth.
Such measurement
will be done in the SNO experiment
by the investigation of the CC process
$ \nu_{e} + d \to e^{-} + p + p $.

\begin{figure}[p]
\begin{center}
\includegraphics[bb=30 280 550 790,width=0.9\textwidth]{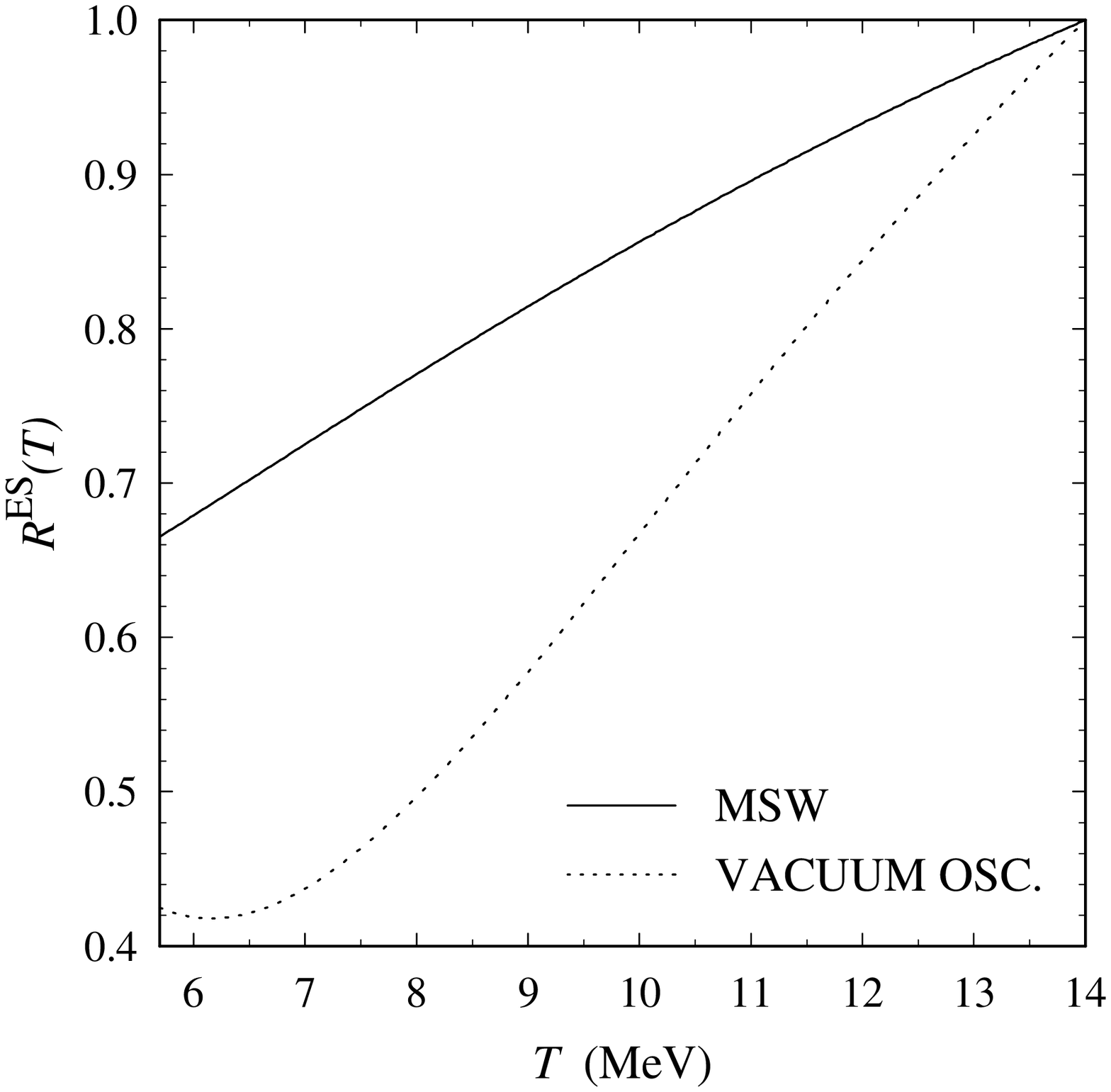}
\\[0.5cm]
Figure \ref{fig5}
\end{center}
\refstepcounter{figure}
\label{fig5}
\end{figure}

If it is found that the function
$ \Sigma^{\mathrm{ES}}(T) / X_{\nu_{\mu}e}(T) $
depends on energy,
we will have a model-independent
proof that solar $\nu_e$'s transfer into sterile states
(if the function
$ \Sigma^{\mathrm{ES}}(T) / X_{\nu_{\mu}e}(T) $
does not depend on
energy,
it could mean either
that solar neutrinos do not transfer into sterile states
or
the probability of this transition does not depend on energy).
The function
$\Sigma^{\mathrm{ES}}(T)$
was calculated in Ref.\cite{BG96}
in a model with $\nu_e$--$\nu_s$ mixing.
The values of the mixing parameters were
taken from the fit of solar neutrino data.
In Fig.\ref{fig5}
we present the function
\begin{equation}
R^{\mathrm{ES}}(T)
\equiv
\frac
{ \Sigma^{\mathrm{ES}}(T) }
{ X_{\nu_{\mu}e}(T) }
\bigg/
\left(
\frac
{ \Sigma^{\mathrm{ES}}(T) }
{ X_{\nu_{\mu}e}(T) }
\right)_{\mathrm{max}}
\;,
\label{res}
\end{equation}
where the subscript max indicates
the maximum value in the allowed range of $T$.
Figure \ref{fig5} illustrates
the rather strong dependence of
the ratio
$R^{\mathrm{ES}}(T)$
on $T$ in the model.

We have described one possible test that could reveal  
the presence of sterile neutrinos in 
the flux of solar neutrinos on the Earth.
Other model-independent tests are discussed in Refs.\cite{BG,BG96}.
If there are no transitions of solar neutrinos
into sterile states,
the initial flux of $^8$B neutrinos
can be determined directly from experimetal data.
From Eq.(\ref{pave}) we have
\begin{equation}
\Phi_{\mathrm{B}}
=
\frac
{ \Sigma^{\mathrm{ES}}(T) }
{ X_{\nu_{\mu}e}(T) }
\;.
\label{phib}
\end{equation}
If the total flux
$\Phi_{\mathrm{B}}$
is known,
the survival probability of solar neutrinos
can be determined from the CC measurement
of the flux of $\nu_e$'s
on the Earth:
\begin{equation}
P^{\mathrm{sun}}_{\nu_{e}\to\nu_{e}}(E)
=
\frac
{ \phi_{\nu_{e}}(E) }
{ X(E) \, \Phi_{\mathrm{B}} }
\;.
\label{peesun}
\end{equation}

\section{Conclusions}
\label{Conclusions}

The present and future neutrino oscillation experiments
will check
the existing indications in favour of neutrino mixing.
We have shown here with different arguments that
these experiments have a good potential to obtain 
model-independent information about
the neutrino mass spectrum and
the elements of the neutrino mixing matrix.
It is clear that this information
will be extremly important
for the future theory of neutrino masses and
mixing.

\begin{flushleft}
\large \bfseries
Acknowledgements
\end{flushleft}

This work was done while one of
authors (S.M.B.) was  
Lady Davis visiting professor at the Technion.
This author would like to thank
the Physics Department of Technion 
for its hospitality.

\end{document}